%% file: main.tex
\documentclass[sigconf]{acmart}
\AtBeginDocument{%
  }

\copyrightyear{2026}
\acmYear{2026}
\setcopyright{cc}
\setcctype{by}
\acmConference[ICSE '26]{2026 IEEE/ACM 48th International Conference on Software Engineering}{April 12--18, 2026}{Rio de Janeiro, Brazil}
\acmBooktitle{2026 IEEE/ACM 48th International Conference on Software Engineering (ICSE '26), April 12--18, 2026, Rio de Janeiro, Brazil}
\acmPrice{}
\acmDOI{10.1145/3744916.3773202}
\acmISBN{979-8-4007-2025-3/2026/04}

\usepackage{cleveref}
\usepackage{xspace}
\usepackage{framed}
\usepackage{subcaption}
\usepackage{xcolor}
\usepackage{listings}
\usepackage{svg}
\usepackage{multirow}

\input{macros}

\begin{document}

\title{Unified Software Engineering Agent as AI Software Engineer}


\author{Leonhard Applis}
\authornote{Joint first authors, ordered alphabetically.}
\email{l.applis@nus.edu.sg}
\orcid{0000-0002-4341-8840}
\affiliation{%
  \institution{National University of Singapore}
  % \city{Singapore}
  \country{Singapore}
}

\author{Yuntong Zhang}
\authornotemark[1]
\authornote{Corresponding author, queries about the paper can be sent to Yuntong Zhang.}
\email{yuntong@comp.nus.edu.sg}
\orcid{0009-0005-1664-7110}
\affiliation{%
  \institution{National University of Singapore}
  % \city{Singapore}
  \country{Singapore}
}

\author{Shanchao Liang}
\email{liang422@purdue.edu}
\orcid{0009-0001-4127-2382}
\affiliation{%
  \institution{Purdue University}
  \streetaddress{Street Address}
  \city{West Lafayette}
  \state{Indiana}
  \postcode{47907}
  \country{USA}
}

\author{Nan Jiang}
\email{jiang719@purdue.edu}
\orcid{0000-0001-8518-2576}
\affiliation{%
  \institution{Purdue University}
  \streetaddress{Street Address}
  \city{West Lafayette}
  \state{Indiana}
  \postcode{47907}
  \country{USA}
}

\author{Lin Tan}
\email{lintan@purdue.edu}
\orcid{0000-0002-6690-8332}
\affiliation{%
  \institution{Purdue University}
  \streetaddress{Street Address}
  \city{West Lafayette}
  \state{Indiana}
  \postcode{47907}
  \country{USA}
}

\author{Abhik Roychoudhury}
\email{abhik@comp.nus.edu.sg}
\orcid{0000-0002-7127-1137}
\affiliation{%
  \institution{National University of Singapore}
  % \city{Singapore}
  \country{Singapore}
}

\renewcommand{\shortauthors}{Applis, et al.}

\begin{abstract}
The growth of Large Language Model (LLM) technology has raised expectations for automated coding. 
However, software engineering is more than coding and is concerned with activities including maintenance and evolution of a project. 
In this context, the concept of LLM {\em agents} has gained traction, 
which utilize LLMs as reasoning engines to invoke external tools autonomously. 
%In an LLM agent, external tools are invoked autonomously in the front-end, while the LLM serves as a back-end. 
But is an LLM agent the same as an AI software engineer? 
In this paper, we seek to understand this question by developing a Unified Software Engineering agent or \useagent. 
Unlike existing work which builds specialized agents for specific software tasks such as testing, debugging, and repair, 
our goal is to build a unified agent which can orchestrate and handle multiple capabilities. 
This gives the agent the promise of handling complex scenarios in software development such as fixing an incomplete patch, adding new features, or taking over code written by others. 
We envision \useagent as the first draft of a future AI Software Engineer which can be a team member in future software development teams involving both AI and humans. 
To evaluate the efficacy of \useagent, we build a Unified Software Engineering bench (\usebench) comprising of myriad tasks such as coding, testing, and patching. 
\usebench is a judicious mixture of tasks from existing benchmarks such as SWE-bench, SWT-bench, and REPOCOD. 
In an evaluation on \usebench consisting of 1,271 repository-level software engineering tasks, \useagent shows improved efficacy compared to existing general agents such as OpenHands CodeActAgent.
For specific tasks such as issue resolution in SWE-bench, \useagent demonstrates efficacy comparable to agents such as AutoCodeRover (which are focused on software maintenance), while maintaining applicability to a broader range of software engineering tasks.
There exist gaps in the capabilities of \useagent for certain coding tasks, 
which provides hints on further developing the AI Software Engineer of the future.
\end{abstract}




\maketitle

% TODO: remove
% \pagestyle{plain} % removes running headers

\input{sections/introduction}

\input{sections/usebench}
\input{sections/background-agents}
\input{sections/useagents}
\input{sections/research-questions-and-setup}
\input{sections/results}
\input{sections/discussion-related-work-threats}

\begin{acks}
This work was partially supported by a Singapore Ministry of Education (MoE) Tier 3 grant ``Automated Program Repair'', MOE-MOET32021-0001, and the NUS Artificial Intelligence Institute (NAII) seed grant NAII-SF-2024-006.
It was also supported in part by NSF 1901242 and 2006688 and a CFI fund.
\end{acks}

\bibliographystyle{ACM-Reference-Format}
\bibliography{references}

\end{document}

%% file: macros.tex
\definecolor{byzantium}{rgb}{0.44, 0.16, 0.39}

\newenvironment{goalbox}[1]{%
  \begin{framed}
  \noindent\textbf{#1.}\hspace{0.5em}%
}{%
  \end{framed}
}

\definecolor{diffstart}{RGB}{0,128,0}
\definecolor{diffincl}{RGB}{0,0,255}
\definecolor{diffrem}{RGB}{255,0,0}
\definecolor{boxbg}{RGB}{240,240,240}

\lstdefinestyle{diffstyle}{
    basicstyle=\ttfamily\footnotesize,
    keywordstyle=\color{blue},
    morecomment=[f][\color{diffstart}]{diff --git},
    morecomment=[f][\color{diffincl}]{+++},
    morecomment=[f][\color{diffrem}]{---},
    numbers=left,
    numberstyle=\tiny\color{gray},
    commentstyle=\color{gray},
    breaklines=true,
    postbreak=\mbox{\textcolor{red}{$\hookrightarrow$}\space},
}

\newcommand{\usebench}{\texttt{USEbench}\xspace}
\newcommand{\useagent}{\texttt{USEagent}\xspace}
\newcommand{\openhands}{\texttt{OpenHands} CodeActAgent\xspace}
\newcommand{\acr}{AutoCodeRover\xspace}
\newcommand{\metaagent}{\texttt{Meta-Agent}\xspace}

\newcommand{\swebench}{SWE-bench\xspace}
\newcommand{\sweverified}{SWE-bench-verified\xspace}

\newif\ifusecolor
\usecolorfalse
\DeclareRobustCommand{\highlight}[1]{%
  \ifusecolor
    \textcolor{blue}{#1}%
  \else
    \ignorespaces#1\unskip
  \fi
}
\newcommand{\highlighttable}{\ifusecolor\color{blue}\else\color{black}\fi}

%% file: sections/introduction.tex
\section{Introduction}
\label{sec:introduction}

Large language models (LLMs) have shown promise in coding, reasoning, and problem solving. 
The model capability itself has improved significantly in terms of reasoning tasks, by training the models to always use reasoning paradigms such as chain of thoughts, without needing further prompting, as shown by GPT o1 \cite{openai2024o1} and its descendant models such as o3. 
On top of improved foundation models, the software engineering community is distilling known best practices 
for downstream tasks into amplifiers for LLMs through agentic systems \cite{yao2023react}. 
An LLM agent for software engineering invokes various interfaces including testing and analysis tools autonomously driven by an LLM reasoning agent. 

While agentic systems inspire researchers \cite{yao2023react,antoniades2024swesearch}, start-ups \cite{moatless_tools,AIDER2024} and  major companies \cite{wadhwa2024masai,rondon2025evaluating},
we currently observe a strong specialization in the emerging technologies among the LLM agents for software. 
Agentless \cite{xia2024agentless} performs program repair, 
Large Language Monkeys \cite{brown2024large} produce unit tests, 
ExecutionAgent \cite{bouzenia2024you} performs a project setup, and so forth. 
These specialized agents are reasonable efforts, especially as they require specialized metrics and datasets. 
However, with many agents comes the need to manage and maintain them as part of the future development environments!

We propose  a unified Software Engineering agent (\useagent) representing a consolidated agentic capability for software engineering tasks. 
Individual SE tasks could be coordinated to result in a more effective ensemble of tasks:
better test-generation should benefit reproduction in program repair, review of patches should lead to better generated code, etc. 
With an integrated software engineering agent, it becomes much more feasible to co-opt it in future development environments \cite{opinion25}.

As a foundation, we need a unified dataset of challenging SE tasks to test our unified agentic capabilities, for which we build Unified Software Engineering Bench, or \usebench for short. 
Recently, the SWE-bench dataset \cite{jimenez2024swebench} has been proposed that captures a set of GitHub issues depicted in natural language and requires bug fixes or feature additions in software projects.  
Thus the SWE-bench dataset is replete with challenges in software maintenance for various real-life software projects. 
\highlight{Our proposed benchmark \usebench is a  meta-benchmark composing a diverse set of software engineering tasks (such as code generation, program repair, test generation) behind a unified application programming interface (API).}
Building a unified API is key, since it allows us to start thinking of a unified agentic capability which can handle different software engineering tasks.
After constructing \usebench, we expand two popular agentic systems, one from industry and one from academia, \acr \cite{zhang2024autocoderover} and \openhands \cite{wang2024openhands, wang2024executable} to solve the tasks in \texttt{USEbench}. 
We consider this a natural evolution or progression of the flurry of research proposing manifold different agents for different software engineering tasks. 
Existing specialized agents usually employ a fixed workflow and approach the given task in a few pre-defined steps.
For example, AutoCodeRover tackles program repair tasks using a fixed two-phase workflow consisting of fault localization and patch generation.
Now, what does it take to generalize an existing program repair agent, with fixed actions and pre-defined workflow, 
so that it can act as a unified software engineering agent (\useagent)? 
For each task or problem in \usebench, the agentic systems are only provided with a description (i.e., a bug report or the documentation of a method to generate) as well as access to a containerized environment of the project. 
As such, our meta-benchmark reveals inherent challenges regarding \textit{task-identification}, \textit{workflow-configuration} and \textit{measuring progress} (esp. determining end-criteria). 

To address these requirements, we equip \acr with a \metaagent, which is instructed to orchestrate the appropriate agents and construct a workflow \textit{on the fly}. 
Over the course of this workflow, the \metaagent utilizes available actions
to construct and maintain a project-state, constituting a structured consensus memory over the trajectories LLM components.
Since \openhands is a general-purpose agent designed to solve a variety of tasks, we consider it for baseline comparison, 
with no architectural changes applied.

We report the efficacy of both agentic systems on \usebench including \texttt{PASS@1} and \texttt{PASS@5} results. 
We perform a detailed error analysis as well as a manual inspection of \textit{positives}, to identify false-positives and cases of over-fitting or memorization. 
On \usebench consisting of 1271 tasks including program repair, regression testing, code generation, and test generation, \useagent achieves an efficacy of 33.3\%, which is higher than the 26.8\% efficacy from the state-of-the-art general agent OpenHands \texttt{CodeActAgent}.
Specifically, on software maintenance tasks in SWE-bench-verified, \useagent has an efficacy of 45.6\% which is similar to the 46.2\% efficacy of the specialized AutoCodeRover agent, while being applicable to more types of tasks.
On test generation tasks that AutoCodeRover could not be applied to, \useagent achieves 31.8\% efficacy, demonstrating its versatility.
We also make the \usebench benchmark public, to encourage further research in design of AI software engineers.

%% file: sections/usebench.tex
\section{Unified Software Engineering Benchmark}
\label{sec:usebench}

\begin{table*}[htb]
    \caption{Summary of primitive and compound tasks unified in \usebench.}
    \centering
    \small
    \begin{tabular}{p{10em} | p{10em} | p{18em} | c | c}
        \toprule
        \textbf{Benchmark} & \textbf{Task} &  \textbf{Concept} & \textbf{Solving Criteria}  & \textbf{\#} \\ \toprule
        \multicolumn{5}{c}{\texttt{Primitive Tasks}} \\ 
        \midrule
        \sweverified & Program Repair & Adjust a program according to an issue. & Test-Suite Pass & 500 \\ \midrule
        SWT-bench-Lite & Regression Testing & Produce a test that asserts a required change, as presented in an issue & Patch-Coverage & 298 \\ \midrule
        REPOCOD-Lite & Code Generation & Generate a method body, matching the functionality outlined in documentation & Test-Suite Pass & 200 \\ \midrule
        REPOTEST-Lite & Test Generation & Test a given method to reach 100\% coverage & Code-Coverage & 173 \\ \midrule
        \multicolumn{5}{c}{\texttt{Compound Tasks}} \\ \midrule
        Partial fix & Test Generation \newline Program Repair & Enrich the SWE-Task with a previously failed, but promising patch & Test-Suite Pass & 100 \\ \midrule
        Feature Development & Code- \& Test-Generation & Add both tests and code, evaluate against the newly produced code coverage & Code-Coverage & 46 \\ 
        \bottomrule
    \end{tabular}
    \label{tab:usebench-combined-overview}
\end{table*}

To make progress towards building a unified agent for software engineering (\useagent), we first build a dataset of automated software engineering tasks. 
There is a rich environment of benchmarks, yet they are somewhat fragmented - they focus on specific types of software engineering tasks.
Popular benchmarks encompass natural language issues for software (\swebench~\cite{jimenez2024swebench}),
automated program repair (Defects4J \cite{just2014defects4j}), 
code generation  (REPOCOD \cite{liang2024can}), 
test generation (SWT \cite{mundler2024swtbench}) 
or documentation generation (CodeNet \cite{CodeNet}), each accompanied by a metric to identify correct or optimal solutions. 
Next to their task, benchmarks often vary in scope too:
the \textit{programming} benchmarks 
(such as HumanEval \cite{HumanEval} , CodeNet \cite{CodeNet}, BigCode \cite{BigCodeProject}, EvalPlus \cite{liu2024your}) cover method-snippets or class-level code that is evaluated against known examples. 
The recently proposed \swebench~\cite{jimenez2024swebench} offers a new angle on repository-scale changes by introducing natural language issues  for program repair instead of a known test-suite. 
However, due to the natural language, some issues were too ambigious or un-solvable --- spurring efforts of OpenAI to create a human-verified subset, called \sweverified~\cite{OpenAISWEBench2023}. 
Solving such natural language issues automatically brings the vision of a future AI software engineer closer to the capabilities of a human software engineer.

In this paper, we propose \usebench, a unified software engineering benchmark as a meta-benchmark which moves beyond project-level code editing tasks suggested by \swebench. 
Our goal is to unify existing individual benchmark sets and design a comprehensive unified dataset capturing multiple SE tasks. 
Our proposal combines a set of existing benchmark sets, namely \sweverified~\cite{OpenAISWEBench2023}, SWT~\cite{mundler2024swtbench}, REPOCOD~\cite{liang2024can} and REPOTEST. 
Out of these \sweverified captures tasks which involve code editing or program modifications to achieve bug fixing / feature addition. 
SWT bench presents validation tasks for testing real-world code fixes in software projects. 
REPOCOD is a recent benchmark which captures code generation tasks in a software project.
REPOTEST is a not previously published derivative of REPOCOD: 
While REPOCOD encompasses repository level code-generation after removing a method body; 
REPOTEST removes tests for a specified method-body and evaluation is based on code-coverage.

The exact composition of \usebench from the constituent benchmarks is shown in 
\Cref{tab:usebench-combined-overview}. 
Six different task types are covered, and for each task type several instances are included in \usebench. 
As a notable technical contribution, we provide a unified interface for agentic interaction with the unified benchmark. 
Previous benchmarks like \swebench provide a suitable interface for evaluation of code editing tasks {\em i.e.}, after providing a \texttt{.json} file tasks are streamlined; still there is significant work involved in interacting with a software project, or extracting data from the constituent files of a software project.
Thus, LLM agents proposed to work on \swebench tasks need to make these adjustments to work with \swebench, e.g. see \cite{moatless_tools}. 
We would want our unified benchmark \usebench to possess a greater degree of usability, so that LLM agents can be evaluated and compared in terms of their efficacy in conducting the software engineering tasks in the benchmark.
To alleviate the burden on researchers, the constituent benchmarks are conveniently plugged behind a common interface revolving around docker images, providing options to read files and execute commands, regardless of the task. 
Due to the unified nature of our benchmark \usebench, we can also imagine unified software engineering agent (\useagent) working on the tasks in \usebench. 
Such a unified agent will be able to achieve complex software engineering tasks, which are solved via a sequence of the software engineering task types covered in the benchmark \usebench. 

We now show two concrete scenarios of more complex software engineering tasks constructed from primitive task types in \usebench, namely \textit{incomplete fix} and \textit{feature development}, elaborated in \Cref{tab:usebench-combined-overview}.
To concretely see how the different task types in \usebench are combined into compound tasks, let us consider the scenario of handling an incomplete code fix. 
In such a scenario,  we (i) test the code and find failing tests, (ii) generate a fix for the failed tests, (iii) test the fixed code and find more failed tests, and (iv) finally generate a fixed code which passes all tests. 
These four steps could be four different runs of a software engineering agent, and could combine different task types from \usebench, namely code fixing and test generation. 
The unified API of tasks in \usebench allows us to combine and alter tasks to capture more complex software engineering challenges, like incomplete fixes.
Another scenario of a complex software engineering task 
is the complete addition of a feature to a codebase. 
This includes a problem description, and requires a functional method-body as well as accommodating tests.
How to solve the task, as well as any form of decomposition, re-iteration or co-evolution is left up to the users of the benchmark.
The final evaluation of this derived scenario will be against a hidden test-suite, as well as the generated tests. 

%% file: sections/background-agents.tex
\section{Background: Agents for Software Engineering}
\label{sec:background-agents}

Agents typically utilize LLMs for decision making and content generation, along with autonomously invoked \textit{tools} to interact with external entities.
These tools allow the agent access to knowledge beyond the LLMs' training data, and allow the agent to influence the external environment through the tool interface.
Agents often employ reasoning frameworks to orchestrate the LLM and the tool usage.
One of the popular reasoning frameworks used in agents is \texttt{ReAct}~\cite{yao2023react}, in which the LLMs are instructed to reason about output-traces and invoke tools in an interleaved manner.

In software engineering, the target system that an agent interacts with is usually a software project. 
Another component of the external system is an execution environment, where software can be executed and tested.
Existing software engineering agents (SE agents) proposed different ways to interact with the software project and the execution environment.
\acr\cite{zhang2024autocoderover} uses a set of project structure-aware tools (e.g. \texttt{search\_method\_in\_class} to navigate the software codebase and gather necessary code context.
It further performs test execution and spectrum-based localization (SBFL)~\cite{wong2016survey} to pinpoint more relevant code locations.
SWE-Agent~\cite{yang2024swe} employs file-based tools to navigate files in the project, for example \texttt{find\_file}, \texttt{open}, \texttt{scroll\_down}, and together they form an agent-computer interface~\cite{yang2024swe}.
OpenHands CodeActAgent~\cite{wang2024openhands, wang2024executable} provides a set of basic tools such as \texttt{CmdRunAction} to execute arbitrary bash commands inside a sandbox environment. 
A similar approach was followed by Google Jules agent~\cite{google2024gemini2}.
Roughly summarized, the existing agents differ (intentionally) in how much structure and domain knowledge they introduce.

Another aspect of agent design is the decision framework.
Existing SE agents have explored options for more controlled decision-making beyond the \texttt{ReAct} framework, in the context of software engineering.
\acr\cite{zhang2024autocoderover} and Agentless~\cite{xia2024agentless} targets software maintenance tasks, and they break the agent's execution into phases like context retrieval/localization and repair.
Another example is RepairAgent~\cite{bouzenia2024repairagent}, which restricts available tools through a finite-state machine.
The different designs in existing SE agents raise an important question -- 
how much \textit{autonomy} should be given to the agent during decision making?
On one hand, agents can follow a fine-tuned workflow for the targeted software engineering (SE) task (e.g. in AutoCodeRover, Agentless, RepairAgent).
In this design, agents operate in fixed phases (e.g. localization is the first phase for program repair),
which makes them more cost-effective~\cite{zhang2024autocoderover,xia2024agentless}.
With the task broken down into phases, it also becomes easier to leverage program analysis techniques in each of the phases.
However, agents with this design have less autonomy, and as a result, they could not be directly applied to another kind of SE task.
Applying a program repair agent to test generation would require drastic changes to its workflow.
On the other hand, agents can be given more autonomy by having access to a general set of tools and the freedom to decide when to use these tools (e.g. in SWE-Agent and OpenHands CodeActAgent).
This general design makes the agents applicable to many kinds of software engineering tasks.
However, this generality makes it harder to exploit domain knowledge and program analysis in the agents.
Moreover, when the task becomes complex and the agent execution gets longer, it is challenging to interpret the agent trajectory consisting of a list of general tool usages.
In comparison, agents operating in ``phases'' are more interpretable, since we can inspect the results of each phase (e.g. what is the identified faulty location at the end of a localization phase).

In this work, our goal is to have an agent applicable to multiple types of software engineering task, 
while still utilizing domain-specific optimizations and interpretability.
We name this agent Unified Software Engineering Agent (\useagent).

%% file: sections/useagents.tex
\section{Unified Software Engineering Agent}
\label{sec:design-of-useagents}
The first step towards a unified software engineering agent is breaking down the fixed workflow of existing agents into components (i.e., phases) and add orchestration to decide which component to invoke based on the task type and \textit{task state}.
To achieve this, we propose a \metaagent, a central LLM-reasoning agent that orchestrates various components, henceforth called \textit{actions}.
The \metaagent and the actions together form a unified software engineering agent (\useagent) that is capable of solving several types of software engineering tasks with a higher level of autonomy.
We identify several challenges in designing such a unified agent to solve tasks in \usebench.

\begin{figure}[t]
    \centering
    \includegraphics[width=0.85\linewidth]{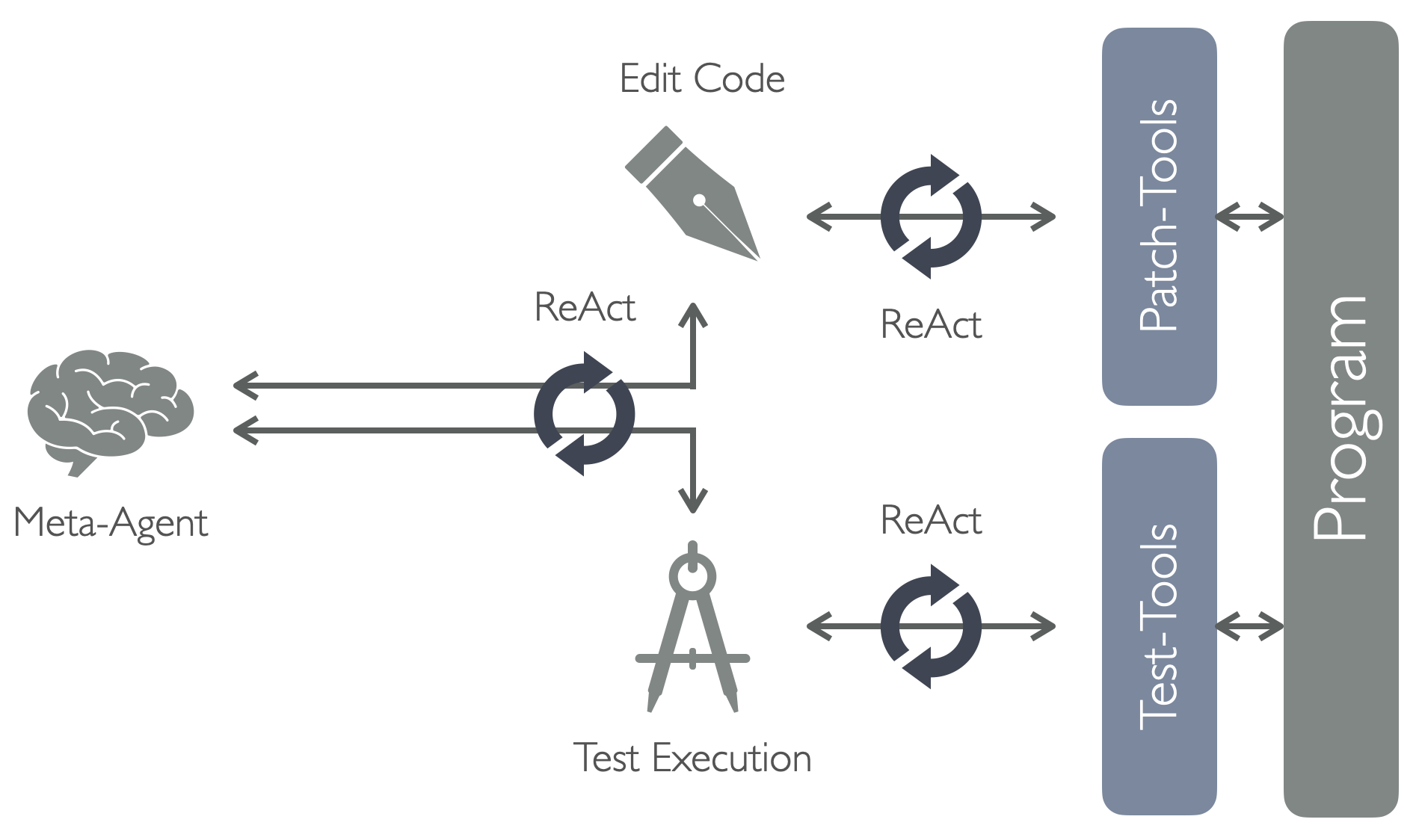}
    \caption{Concept: Meta-Agent abstracting over actions.}
    \label{fig:react-hierarchy}
\end{figure}

The first challenge is to adapt to tasks beyond a fixed workflow. 
Previous agents extract actions such as context retrieval or patch generation as clear ``units of work'' from developer workflows and organize them into state-machines or fixed workflows.
This rigid organization of actions restricts their adaptability to other tasks.
To achieve high agent autonomy we allow the actions to be freely composed and
introduce a \metaagent that orchestrates the actions using a \texttt{ReAct}-style loop, forming a \textit{meta-layer} over the actions, shown in \Cref{fig:react-hierarchy}. 
Starting from natural-language descriptions, the \metaagent invokes actions, observes the changes made by the action, reasons about the output, and decides on the next action to take.
This framework enables \textit{on-the-fly} composition of workflows, adjusting to different tasks, errors and changing states.
\texttt{ReAct}-style decision making has been widely used in previous agents to orchestrate \textit{tools} that directly interact with the codebase; which we extend towards \textit{action} orchestration.
The overall trajectory and choices of the \metaagent will construct an \textit{action-graph}.
In our action-graph, any node can be the start or end-point - for example, depending on the task, an action of ``executing tests'' can be the starting point or the last assurance before termination.
The agent execution can be seen as an implicit construction of this action-graph on the fly.

The next challenge is to provide the \metaagent with a set of reliable, yet flexible actions that are suitable for composition.
Existing general agents such as OpenHands \texttt{CodeActAgent}~\cite{wang2024openhands} and Google Jules~\cite{google2024gemini2} provide the LLM with a console to execute \textit{any} command.
The action of executing a console command is extremely flexible, but hard to control.
The agent's execution consists of a sequence of console commands, which are difficult for human developers to interpret and raise security concerns.
Thus, we propose to design our actions at a coarser granularity, where each action encapsulates a ``unit of work'' that developers carry out during the software engineering process.
Each action receives instructions from the \metaagent on \textit{what outcomes it is supposed to achieve}, rather than \textit{how it achieves them}.
The \metaagent and individual actions communicate through this intent-based interface.
An example action is \texttt{EditCode}, where the \metaagent specifies what should be edited in the code, but not the exact details of making the edits.
This design results in a set of modularized actions with clear responsibilities, that can be tested, extended and maintained individually. 
Moreover, when one action is improved, the improvements propagate across all involved task types. 
We present a sample list of actions for unified software engineering agents in Section~\ref{subsec:instantiating}.

The third challenge is knowledge management: 
For complex tasks, the information provided \textit{a priori} is not sufficient, 
and a central effort in solving tasks is identifying and producing relevant information. 
Once produced, this context should be made available to the agentic system, i.e. \metaagent and other actions.
Existing work has defined this problem as \textit{memory management} of agents~\cite{zhang2024survey, han2024llm}, where memory refers to historical information relevant to the current task.
In literature on LLM agents, memory is often categorized into short-term (e.g., information within an ongoing conversation), long-term (e.g., historical data from previous sessions), and consensus memory (e.g., shared knowledge between multiple agents)~\cite{han2024llm}.
In this work, we explicitly identify what information to memorize when designing a unified software engineering agent, and classify them into the different types of memory.
We categorize the results of actions as short-term memory, which is only presented to the \metaagent to decide the next action.
Project-specific knowledge, such as developer-written documentation, is considered long-term memory.
These documents are embedded and stored in a vector database 
accessible for Retrieval-Augmented Generation (RAG) by the agent.
Finally, artifacts generated by actions that are useful in guiding other actions constitute consensus memory.
Such artifacts include program locations or code edits that can be notable to other actions.
We introduce a structured \textit{task state} that represents the consensus memory, which can be read and modified by different actions.

\subsection{Instantiating \acr as \useagent}
\label{subsec:instantiating}
In this section, we highlight steps to build the unified SE agent \useagent that handles multiple types of SE tasks.
We design our \useagent drawing inspiration from the open-source agent 
\acr~\cite{zhang2024autocoderover,ruan2024specrover}, since \acr operates in different phases which are suitable as starting points for actions.
We elaborate the design of \useagent and flesh out our proposed solutions to the challenges in \Cref{sec:design-of-useagents}.  

\begin{figure}[t]
    \centering
    \includegraphics[width=0.9\linewidth]{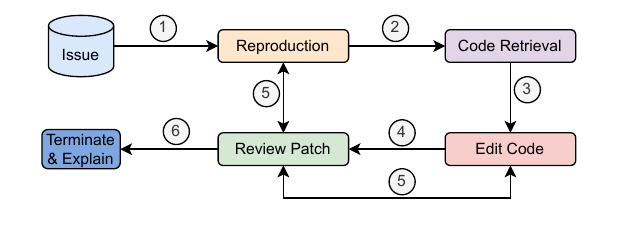}
    \caption{Overview of \acr. \acr composes a fixed workflow for program maintenance tasks.}
    \label{fig:overview-acr-flow}
\end{figure}

\acr is an LLM agent that targets program maintenance tasks. 
Figure~\ref{fig:overview-acr-flow} shows the overall workflow of \acr.
Starting from a natural-language description of a software issue, \acr first writes an executable self-contained test that reproduces the issue (Step \textcircled{1}).
The stacktrace of the reproducer test, together with the task description, is sent to a context retrieval component to identify relevant code locations (Step \textcircled{2}).
The relevant program locations are made available to the \textit{Edit Code} component for writing a candidate patch (Step \textcircled{4}).
With both LLM-generated patch and reproducer test, the \textit{Review Patch} component then executes the tests on the patched program, decides whether the are correctly written, and iteratively improves them (Step \textcircled{5}).
Finally, the \acr workflow finishes when the reviewer approves a patch or the execution reaches a pre-defined round limit (Step \textcircled{6}).
\acr proved effective for resolving software issues such as bug fixing~\cite{zhang2024autocoderover,ruan2024specrover}.
However, these fixed transitions among components make agents like \acr not generalizable to task types that require a different workflow.
For example, in a task like test-suite generation, steps such as Reproduction and Review Patch are irrelevant, requiring the design of a new workflow or even a new agent.

\begin{figure}[t]
    \centering
    \includegraphics[width=0.95\linewidth]{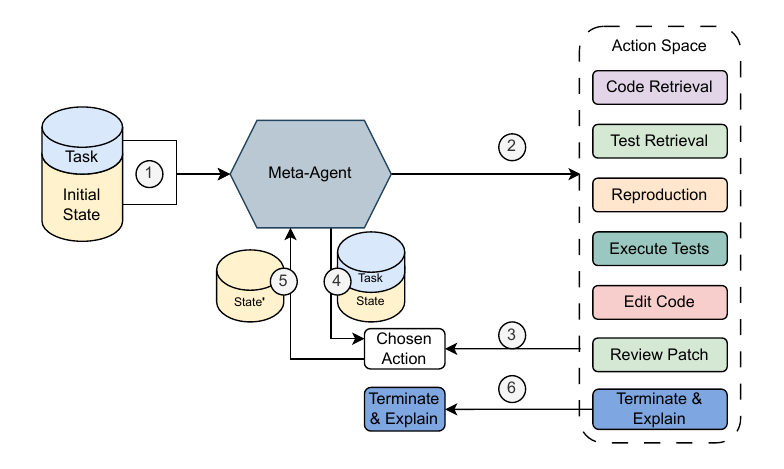}
    \caption{Overview of the \useagent and workflow. The Meta Agent chooses available actions, provides the state and retrieves a altered state until termination is chosen.}
    \label{fig:overview-useagent-flow}
\end{figure}

Instead of designing a new agent for each task type, our goal is to build \textit{one} general agent that handles multiple task types.
To achieve this, we disassemble the workflow of \acr and reassemble it into \useagent shown in Figure~\ref{fig:overview-useagent-flow}.
We adapt components in \acr such as Code Retrieval into \textit{actions} that can be freely composed by a \metaagent for different task types.
Given the task description that mentions the task type in natural language, the \metaagent uses \texttt{ReAct}-style reasoning to select next actions to execute.
Short-term feedback such as action execution results are reflected directly to the \metaagent, while artifacts that should be preserved longer are stored into a \textit{task state} representing the consensus memory among actions.
We next discuss each component of \useagent in more details.

\begin{table*}[h]
    \caption{\centering Details of actions used in \useagent. \texttt{Diff} denotes a Patch to the project, and subscript specifies modifiers. \newline E.g. \texttt{Diff\textsubscript{Code}} expresses a change to program code, instead of test code.}
    \centering
    \small
    \begin{tabular}{c|p{13em}|p{16.5em}|p{12em}}
        \toprule
        
        \textbf{Action} & \textbf{Description} & \textbf{Action Signature} & \textbf{Interaction with Task State} \\
        
        \hline
        
        CodeRetrieval  & Collect program code relevant to the task description.  & \texttt{Description} $\rightarrow$ \texttt{Locations\textsubscript{Code}} & \texttt{Write:} \texttt{Locations\textsubscript{Code}}  \\ 
        
        \hline
        
        TestRetrieval & Collect test code relevant to the task description.  & \texttt{Description} $\rightarrow$ \texttt{Locations\textsubscript{Tests}} & \texttt{Read:} \texttt{Locations\textsubscript{Code}} \newline \texttt{Write:} \texttt{Locations\textsubscript{Test}}  \\ 
        
        \hline
        
        Reproduction  & Generate a system-level standalone reproduction test.  & \texttt{Description} $\rightarrow$ \texttt{ Diff\textsubscript{Test}} & \texttt{Write:} \texttt{Test\textsubscript{Reprod.}} \newline \texttt{Write:} \texttt{Output(Test\textsubscript{Reprod.})}\\ 
        
        \hline
        
        ExecuteTests  & Execute a subset of the program test suite.  & \texttt{List[TestFile]} $\rightarrow$ \texttt{Summary(Results)} & \texttt{Read:} \texttt{Locations\textsubscript{Test}} \newline \texttt{Write:} \texttt{Command\textsubscript{Test}}\\ 
        
        \hline
        
        EditCode & Make modifications across the codebase, including test-code.  & $\Delta$\texttt{Behavior} $\times$ \texttt{Location} $\rightarrow$ \texttt{Diff} & \texttt{Read:} \texttt{Locations\textsubscript{Code}} \newline \texttt{Read:} \texttt{Locations\textsubscript{Test}} \newline \texttt{Write:} \texttt{Diff} \\ 
        
        \hline
        
        ReviewPatch  & Review and iteratively improve a generated patch
and generated test.  & \texttt{Test\textsubscript{Reprod.}}$\times$\texttt{Diff\textsubscript{Code}} $\rightarrow$ \texttt{Bool} & \texttt{Read:} \texttt{Diff\textsubscript{Code}} \newline \texttt{Read:} \texttt{Test\textsubscript{Reprod.}} \newline \texttt{Write:} \texttt{Diff\textsubscript{Code}}  \\ 

        \hline

        Terminate  & Finish the agent execution and select a final solution.  & \texttt{State} $\rightarrow$  \texttt{Diff} & -  \\
        
        \bottomrule
    \end{tabular}
    \label{tab:useagent-actions}
\end{table*}

\paragraph{Actions.} Given the tasks of program repair, regression testing, code generation, and test generation in \usebench, we identify a priming set of actions for \useagent. 
The set of actions are shown in~\Cref{tab:useagent-actions}.
Each action provides the \metaagent with an interface that specifies its description, input, and output, but abstracts away the underlying execution details.
In addition to interacting with the \metaagent through input and output, each action can also read from and write to the task state.

Many of the actions in~\Cref{tab:useagent-actions} are conceptually inspired by those used in existing agents such as \acr (Figure~\ref{fig:overview-acr-flow}). 
On top of the existing agents, we designed a few new actions that are essential for the task types in \usebench.
These new actions include \texttt{TestRetrieval} and \texttt{ExecuteTests}.
The \texttt{TestRetrieval} action explores the codebase and retrieves testcases relevant to the current task.
It works similarly compared to the existing \texttt{CodeRetrieval}, which employs a set of search tools such as \texttt{search\_func} to search for relevant code units in the program.
The \texttt{ExecuteTests} action executes parts of the project test suite, which is useful for validating code/test changes by execution.
Current agents focusing on SWE-bench usually assume the commands to run existing unit tests are given as an input to the agent~\cite{ruan2024specrover,xia2024agentless}; however, this assumption requires additional manual configuration when setting up the agents on new projects.
Our \texttt{ExecuteTests} action does not assume the test commands to be given, and instead it queries the project documentation files to retrieve the project-specific commands through RAG.
In addition, we added a special \texttt{Finish} action that signals the termination of the agent execution. 
The \texttt{Finish} action is invoked with an argument specifying the final result for the current task.
Unlike other actions which encapsulates concrete workflows, the \texttt{Finish} action only terminates the agent and outputs its argument as the final result from the agent.

These actions can be invoked multiple times and combined in various ways to complete a given task.
For example, for a code generation task, the \metaagent may first invoke \texttt{CodeRetrieval} to understand the surrounding code context, invoke \texttt{EditCode} to draft the implementation, invoke \texttt{TestRetrieval} to find out where existing tests for this code component resides, and then invoke \texttt{EditCode} to write new tests for the newly added code. 
It may then enter a refinement loop of improving the new code and tests using the \texttt{ExecuteTests} and \texttt{EditCode} actions.
We note that this sequence of action invocation can happen without prior configuration, as we will show in \cref{subsec:results-rq3}.

\paragraph{Task State.} Our next contribution towards a \useagent is the design of a \textit{task state}.
The task state represents the \textit{consensus} memory among the actions - each action can write new artifacts generated by them into the task state so that the artifacts can be visible to other actions.
Our task state captures essential artifacts involved in a software debugging process, including relevant locations, test execution information, and prior attempts of code modifications.
Specifically, the task state $S$ is defined as: 
\(
S=(L_c,L_t,R_{exec},DS)
\).
Here, $L_c$ represents the relevant code locations (e.g., program methods and classes); 
$L_t$ represents the test locations (e.g., unit test methods).
The locations are identified by the retrieval actions, and are utilized by downstream code editing actions.
$R_{exec}$ contains the execution results of different 
tests against different candidate patches.
These execution information provides guidance to other actions in improving partial solutions and selecting the final solution.
The last part of the task state is a \textit{diff store} ($DS$), which records all the generated diff contents from the \texttt{EditCode}, \texttt{Reproduction} and \texttt{ReviewPatch} actions.
These diff contents include modifications to both the program code and the test code.
The diff store is necessary to allow for versatile invocations of the \texttt{ExecuteTests} and \texttt{EditCode} actions.
By selecting different subsets of diffs in $DS$ and passing them to \texttt{ExecuteTests}, tests can be executed on different versions of the project after applying the selected diffs.
Similarly, \texttt{EditCode} can first apply an existing patch from $DS$, and subsequently introduce additional modifications, thereby enhancing partial patches.
\highlight{Concretely, when an action is implemented, the implementation defines which part of the action output will be stored into the task state.}

\paragraph{MetaAgent.} 
The last important addition is the \metaagent which orchestrates the various actions.
As shown in Figure~\ref{fig:overview-useagent-flow}, \metaagent takes in the task description and an initial empty task state (Step \textcircled{1}), and then iteratively selects an action from the action space to execute (Step \textcircled{2}).
Each action represents an encapsulated workflow to complete a ``unit of work'', and exposes its input/output interface to the \metaagent.
At each step of action selection, we present the \metaagent with the current task state, the overall task description, as well as the output from the previously invoked action.
The invoked action modifies the task state and retunrs it back to the \metaagent (Step \textcircled{5}).
This interleaving of reasoning (i.e. select the next action based on current output and state) and action (i.e. execute the actual action to obtain new observations) forms a \texttt{ReAct}-style loop at the action level.
When the \metaagent deems that one of the diffs in the diff store is a satisfactory solution to the given task, it invokes the \texttt{Finish} action to end the agent execution and output the final solution.
In the case where the \metaagent could not decide on the final solution (i.e. could not invoke the \texttt{Finish} action) after a pre-defined limit for the ReAct-loop, we ends the execution and invoke the LLM to select one of the candidates in the diff store as the final solution.

\subsection{Configuring OpenHands}
OpenHands \texttt{CodeActAgent} \cite{wang2024openhands} is a general-purpose agent designed to solve general tasks\footnote{Note: OpenHands is also developing \texttt{CodeActSWEAgent}, specialised on SWE-Bench. We purposefully chose the non-specialised implementation for \usebench. The entries on the SWE-Bench leaderboards are to our knowledge results from \texttt{CodeActSWEAgent}.}. 
It utilizes a structured controller-agent-runtime architecture, where the \texttt{AgentController} acts as a supervisor, enforcing operational constraints (such as conversation iterations and budget) and managing the agent's lifecycle (start, stop, pause). 
The \texttt{CodeActAgent} makes decisions, interprets LLM responses, and converts them into actions to interact with the runtime, which is an isolated sandbox environment.
Since \texttt{CodeActAgent} does not rely on a predefined task-specific workflow, it can be applied to a wide range of software engineering tasks, including those in \usebench, without requiring modifications to its architecture. 

During execution, OpenHands \texttt{CodeActAgent} operates iteratively, generating actions, executing them, and processing observations before determining the next step. 
The actions include:
\setlength{\leftmargini}{0pt}
\begin{itemize}
    \item \texttt{CmdRunAction}: execute Linux bash commands.
    \item \texttt{IPythonRunCellAction}: execute Python code in Jupyter or IPython environment.
    \item \texttt{FileEditAction}: read, write, or edit files in the runtime.
    \item \texttt{AgentFinishAction}: signal task completion or termination.
\end{itemize}
The execution of \texttt{CodeActAgent} terminates when the agent issues an AgentFinishAction, reaches resource limits (e.g., maximum iterations), or encounters an error. 
In practice, we design different prompt templates for each task type as input to OpenHands \texttt{CodeActAgent}, incorporating both the task type and task description.
Compared to \useagent, the OpenHands \texttt{CodeActAgent} employs a similar reasoning framework, but the actions operate on a lower-level such as execution of a single bash command.
It follows a single \texttt{ReAct}-loop centered around a more open set of available actions. 
This free-wheeling approach is an alternative to the more structured and pronounced options in \Cref{subsec:instantiating}.
\highlight{
These two designs demonstrate a trade-off: allowing arbitrary bash commands makes the agent more flexible, while using structured actions improves interpretability and control over the agent.
}

%% file: sections/research-questions-and-setup.tex
\section{Research Questions \& Experiment Setup}
\label{sec:rqs}

Our first research question focuses on 
evaluating and comparing agentic systems for tasks beyond program repair. 

\begin{goalbox}{\textbf{RQ1: Efficacy of Agentic Systems}}
    How well do state-of-the-art agentic systems perform across the diverse software engineering tasks in \usebench?
\end{goalbox}

We address \textbf{RQ1} by applying \useagent and OpenHands to all data points outlined in \Cref{sec:usebench} and report a \texttt{PASS@1}. 
In addition to the efficacy, we also report an investigation of the major obstacles for the individual benchmarks.
To understand the effect of randomness, we report \texttt{PASS@5} on a significant subset of the data points. 
\highlight{
We also examine the feasibility of using open-source models as the backend of \useagent, by evaluating \useagent with DeepSeek-V3.
% Similarly, to compare the feasibility of open source models, we investigate the efficacy of Deepseek-V3 on the significant subset.
}

A known issue and motivation for our next research question is overfitting in plausible patches. 
Such patches pass the evaluation-criteria by bypassing the harness with \textit{tricks} rather than benign functionality. 
Sometimes, actual solutions are checks for edge-cases - these can only be distinguished from overfitting by consulting the ground truth (i.e., the gold patch). 
We aim to identify and quantify overfitting solutions.

\begin{goalbox}{\textbf{RQ2: Analysis of Plausible Patches}}
    What is the rate of overfitting among the \textit{resolved} patches?
\end{goalbox}

We sample a significant subset\footnote{at 95\% confidence with 10\% error margin.} of solved data points and manually examine whether the plausible solution is overfitting.
During this manual investigation, we also look for \textit{memorization} in plausible patches, i.e. patches that perfectly mirror the ground truth developer patch or contain suspicious content. 

As described in \Cref{sec:design-of-useagents}, one milestone is to make the agent adapt autonomously to different tasks. 
We study this adaptation by investigating the resulting sequence of actions in \texttt{RQ3:}

\begin{goalbox}{\textbf{RQ3: Effectiveness of Self-Configuration \highlight{and Actions}}}
    \highlight{Can \useagent self-configure based on the task type while maintaining reasonable efficacy? How does adding new actions affect efficacy?}
\end{goalbox}

We present the sequence of actions of the \useagent and accumulate their patterns per task. 
We showcase examples on how these top-level patterns translate into commands, queries and edits. 
\highlight{Furthermore, to quantitatively understand the effect of using dynamic decision making, we perform an ablation study by using a static action sequence for each task type. We also perform an ablation study by removing one of the newly added actions \texttt{ExecuteTests}.}

Lastly, we identify open challenges in the current unified software engineering agent. 
We formulate key difficulties and discuss potential future directions to address them.

\begin{goalbox}{\textbf{RQ4: Open Challenges for Agentic Systems}}
    What are the challenges faced by the current unified agent in tasks in \useagent?
    Can we identify actions to remediate them? 
\end{goalbox}

\paragraph{Experiment setup.} 
We evaluate agentic systems such as \useagent and OpenHands \texttt{CodeActAgent} on the full \usebench which consists of 1271 datapoints.
In most of our experiments, agentic systems use Anthropic Claude 3.5 Sonnet v2 (claude-3-5-sonnet-20241022) as the backend LLM. 
\highlight{
Additional experiments with DeepSeek-V3 invokes the DeepSeek-V3 model through the Openrouter API \footnote{\url{https://openrouter.ai/deepseek/deepseek-chat-v3-0324}}}.
For \useagent, we set the maximum rounds of action invocation from the \metaagent to 20, and force the agent to end execution and select a candidate solution. 
We set the model temperature to zero in \useagent.
To better understand the effect of randomness, we randomly sampled a statistically significant subset (295 instances) to report \texttt{PASS@5} results from both agentic systems.

%% file: sections/results.tex
\section{Results}
\label{sec:results}

\subsection{RQ1: Efficacy of Agentic Systems}

An overview of the efficacy of the agents on \usebench are shown in ~\Cref{tab:efficacy}.
We investigate the results per system.

\newcommand{\tinyclaude}{\raisebox{-0.3ex}{\includegraphics[height=1em]{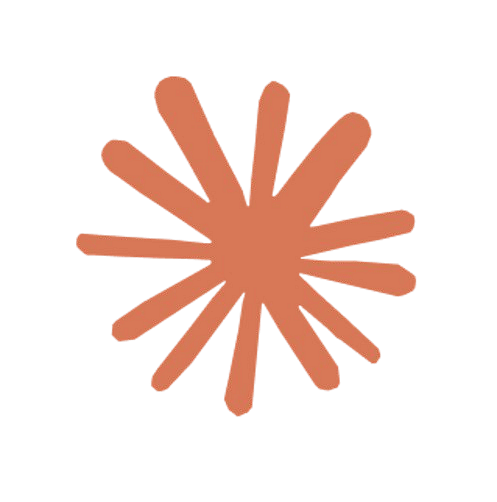}}}
\newcommand{\tinydeepseek}{\raisebox{-0.3ex}{\includegraphics[height=1em]{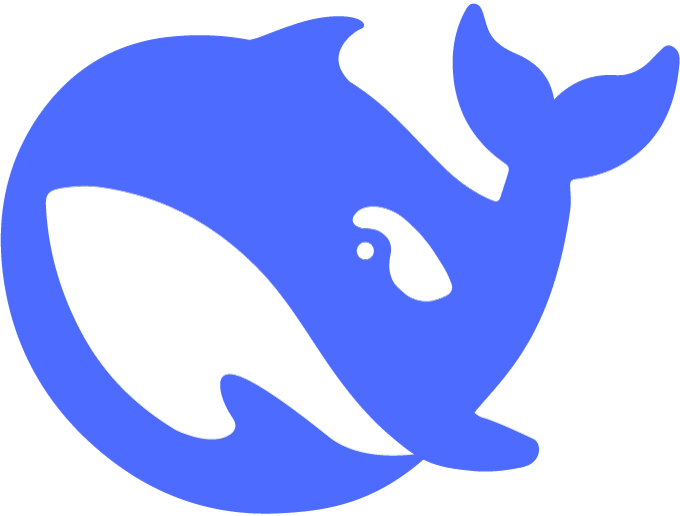}}}

\begin{table*}[t]
    \centering
    \footnotesize
    \caption{Efficacy of \useagent and OpenHands \texttt{CodeActAgent} on \usebench.}
    \label{tab:efficacy}
    \setlength{\tabcolsep}{4pt}
    \begin{tabular}{lc|ccc|ccc}
        \toprule
        &  & \multicolumn{3}{c|}{\useagent} & \multicolumn{3}{c}{OpenHands} \\
        \cmidrule(lr){3-5} \cmidrule(lr){6-8}
        Dataset & Tasks & \tinyclaude PASS@1 & \tinyclaude\footnotemark[4]
 PASS@5 & \tinydeepseek\footnotemark[4]
 \highlight{PASS@1} & \tinyclaude PASS@1 & \tinyclaude\footnotemark[4]
 PASS@5 & \tinydeepseek\footnotemark[4]
\highlight{PASS@1} \\
        \midrule
        SWE-Ver       & 500  & 228 (45.6\%)  & 66.7\%  & \highlight{42.7\%} & 192 (38.4\%)  & 52.1\% & \highlight{35.9\%} \\
        SWT           & 298  & 120 (40.3\%)  & 53.6\%  & \highlight{34.8\%} & 85 (28.4\%)   & 50.7\% & \highlight{29.0\%} \\
        REPOCOD       & 200  & 12 (6.0\%)    & 15.2\%  & \highlight{6.5\%}  & 11 (5.5\%)    & 6.5\%  & \highlight{0\%} \\
        REPOTEST      & 173  & 55 (31.8\%)   & 42.5\%  & \highlight{17.5\%} & 45 (26.0\%)   & 42.5\% & \highlight{10.0\%} \\
        \midrule
        SWETRY        & 100  & 8 (8.0\%)     & 30.4\%  & \highlight{8.7\%}  & 7 (7.0\%)     & 8.7\%  & \highlight{4.3\%} \\
        \midrule
        Total         & 1271 & \textbf{423 (33.3\%)}  & 49.5\% & \highlight{29.1\%} & \textbf{340 (26.8\%)} & 44.1\% & \highlight{22.6\%} \\
        \bottomrule
    \end{tabular}
\end{table*}

\paragraph{\useagent.}

In general, \useagent demonstrates its applicability across different types of tasks in \usebench.
\useagent achieves a \texttt{PASS@1} of \textbf{45.6\%} on \textbf{SWE-Verified}. 
In comparison, \acr, a more specialized agent on issue resolution, achieved efficacy of \textbf{46.2\%} on the same SWE-Verified benchmark when using the same Claude 3.5 Sonnet v2 backend LLM~\cite{SWE-bench-leaderboard}.
\useagent relaxed the fixed workflow in \acr and is a more general agent that applies to more task types.
Although being more general, \useagent demonstrates similar efficacy on the specialized issue resolution task compared to AutoCodeRover.
This comparable performance demonstrates that, with the design of \useagent, the increased autonomy does not result in noticeable reduction in efficacy.

While being effective on issue resolution tasks, \useagent can handle other tasks which specialized agents could not be applied to.
For issue reproduction tasks in \textbf{SWT}, \useagent resolved \textbf{40.3\%} of the tasks.
Here, ``resolved'' means \useagent generated test cases that cover all changed lines in the developer-written patch for the issue, without seeing the patch. 
Moreover, for a significant number of unresolved tasks, \useagent generated test cases that achieve a coverage above 90\%.
In these cases, the generated tests do not fully satisfy the requirement of the benchmark, but only miss coverage on individual statements or branches. 
On test generation tasks in \textbf{REPOTEST}, \useagent achieves efficacy of \textbf{31.8\%}, which is slightly lower than similar tasks in SWT.
While the requirement in SWT is to cover a developer-written patch, the requirement in REPOTEST is to generate tests that cover an entire method, which can be more challenging.
Sometimes the agent generates tests that cover most scenarios in a method, but do not cover all the lines.
On the other hand, we observe certain advantages of agent-generated tests compared to developer-written tests.
The agent often generates shorter tests where each test has individual assertions, 
while the developer-written tests tend to consist of long tests that combine multiple inputs and assertions.
In a real test-suite, short individual tests can provide more informative error messages when part of the test-suite fail \cite{spadini2018relation,van2001refactoring}.

Tasks in \textbf{REPOCOD} proved to be the most challenging with only \textbf{6\%} resolution rate. 
REPOCOD tasks are generally challenging because the requirement is to generate a complete method implementation that can pass a large number of hidden test cases.
Overall, for many unresolved instances, the generated method passes a high number of tests, yet does not account for a few edge-cases.
The majority of resolved instances stem from Sympy, a library for symbolic mathematics. 
We expect a relation between the relative success in Sympy and the fact that mathematical reasoning is becoming a common aspect of training in foundational models.

On the compound benchmark \textbf{SWETRY}, \useagent has a resolve rate of \textbf{8\%}, which shows some promise to apply \useagent as a follow up to partial fixes. 
We need to stress that the data points in SWETRY are sourced from previous failed attempts of agentic systems, which means they are from the more challenging instances in SWE-verified.
We observe two main failure modes on the SWETRY tasks.
Firstly, since the partial patch is given to the agent as part of the task description, the agent often attempts to derive a solution on top of this partial patches by making modifications to it. 
However, the partial patch may be misplaced.
In these cases, the agent still chose to iterate over this patch, and failed to discard the partial attempt and approach the task in a different way.
Secondly, the \metaagent sometimes ends the agent execution prematurely.
For example, when the \texttt{ExecuteTests} action still shows some relevant test errors, the \metaagent wrongly disregards errors as irrelevant to the task, ending the workflow.
This can be addressed by configuring the prompt, at the cost of longer execution time.

\paragraph{\texttt{OpenHands}.}
OpenHands \texttt{CodeActAgent} achieves a \texttt{PASS@1} of \textbf{38.4\%} on \textbf{SWE-verified}, which is lower than the reported result from the SWE-Bench leaderboard~\cite{SWE-bench-leaderboard}.
We attribute this difference to the choice of agent - the reported result on the leaderboard is generated using a specialized \texttt{CodeActSWEAgent}, while we used the more general \texttt{CodeActAgent} in our evaluation.

We utilize the general \texttt{CodeActAgent} to solve the different types of tasks in \usebench. 
For \textbf{SWT}, \texttt{CodeActAgent} achieves a resolution rate of a \textbf{28.4\%}, slightly lagging behind \useagent. 
Similar to that of \useagent, \texttt{CodeActAgent} only achieves a low resolution rate on \textbf{REPOCOD} (5.5\%). However, unlike \useagent, resolved instances are evenly distributed across \texttt{sympy}, \texttt{astropy} (a popular library for Astronomy and Astrophysics), and \texttt{Plotly} (a graphing library), each contributing 27\% of the total resolved cases.
On \textbf{REPOTEST}, OpenHands \texttt{CodeActAgent} resolves 26.0\% of the total instances, which is similar but slightly lower compared to its performance on SWT. 
This trend is consistent with that of \useagent. On \textbf{SWETRY}, \texttt{CodeActAgent} only solves 7\%, similar to that of \useagent, indicating that 
improving partial fixes is still a difficult task for general agents.

\paragraph{\highlight{Open-Source Model}}
\highlight{
We also report the efficacy of the agents with the open-source model DeepSeek-V3 in Table~\ref{tab:efficacy}.
With DeepSeek-V3, both \useagent and OpenHands have a slight performance drop, which is expected. 
% There is a slight performance drop when switching to an open-source model, which is expected. 
There is a larger performance drop on REPOTEST for both agents. 
REPOTEST requires generating a large amount of test code to achieve full test coverage -- DeepSeek-V3 sometimes makes minor mistakes in the generated code, or generates tests that only partially cover the target function.  
}

\paragraph{PASS@5}
Moving to a \texttt{PASS@5} increases the efficacy of \useagent to 49.5\% (\textbf{+48.6\%}) and \texttt{OpenHands} to 44.1\% (\textbf{+64.6\%}). 
This shows a similar increase across the two systems when employing retries.

\footnotetext[4]{\highlight{Derived from the statistically significant subset.}}

\begin{goalbox}{\textbf{Summary: RQ1 - Efficacy of Agentic Systems}}
    \useagent solves the tasks from \usebench at a 33.3\% rate, compared to 26.8\% of \texttt{OpenHands}. 
    There is a consistent improvement over all benchmarks. 
    \useagent's efficacy of 45.6\% for SWE-verified is close to the performance of fine-tuned systems.
\end{goalbox}

\subsection{RQ2: Dissection of Plausible Patches}
\label{subsec:plausible-patch-analysis}

%% Numbers used for RQ3 - Overfitting and Patch Analysis
\newcommand{\numplausiblesamplesize}{218\xspace}
\newcommand{\numplausibleoverfit}{23\xspace}
\newcommand{\percentplausibleoverfit}{10.5\%\xspace}
\newcommand{\percentsweplausibleoverfit}{13.6\%\xspace}
\newcommand{\numplausiblememorize}{3\xspace}
\newcommand{\percentplausiblememorize}{1.3\%\xspace}
\newcommand{\numplausibleanomaly}{33\xspace}
\newcommand{\percentplausibleanomaly}{15.1\%\xspace}

For \useagent, we manually investigated a sample of \numplausiblesamplesize plausible solutions (i.e. passing the resolution criteria of each benchmark) to understand the degree of memorization, overfitting or other anomalies. 
We report a low number of \numplausiblememorize cases of memorization (\textbf{\percentplausiblememorize}), next to \numplausibleoverfit cases of overfitting (\textbf{\percentplausibleoverfit}). 
The percentage of overfitting solutions is lower than the previously reported overfitting rate of 31\% on SWE-bench from SpecRover~\cite{ruan2024specrover}.
We attribute the lower overfitting rate in part to our extensive \texttt{ExecuteTests} action in \useagent, which allows for more versatile test execution.
As a result, we see an overfitting rate of 13.6\% from \useagent on the SWE-verified dataset.
In addition, the SWT and REPOTEST datasets show low numbers of overfitting, which further reduces the overall rate.
Furthermore, we identified a set of \numplausibleanomaly anomalies that are plausible, and do not overfit, yet they are different from the ground truth. 
One example anomaly includes the introduction of a new vector library for \texttt{xarray} (a vector library itself).
Another example is an unorthodox class-inheritance check for \texttt{scikit-learn} - instead of utilizing pythons standard function \texttt{isinstance}, 
a predicate is applied to the objects \texttt{obj.\_\_classes\_\_}. 
These solutions are functionally correct but are unlikely to be accepted by developers, so we have marked them as anomalies.

\begin{goalbox}{\textbf{Summary: RQ2 - Analysis of Plausible Solutions}}
    We manually inspected \numplausiblesamplesize plausible solutions from \useagent and report \percentplausibleoverfit overfitting, 
    \percentplausiblememorize memorization. 
    Overfitting was most common in \texttt{SWE-verified}, and least common in \texttt{REPOTEST} and \texttt{SWT}. 
\end{goalbox}

\subsection{RQ3: Self-Configuration}
\label{subsec:results-rq3}

\begin{figure*}[t]
    \centering
    \begin{subfigure}{0.49\linewidth}
        \centering
        \includegraphics[width=\linewidth]{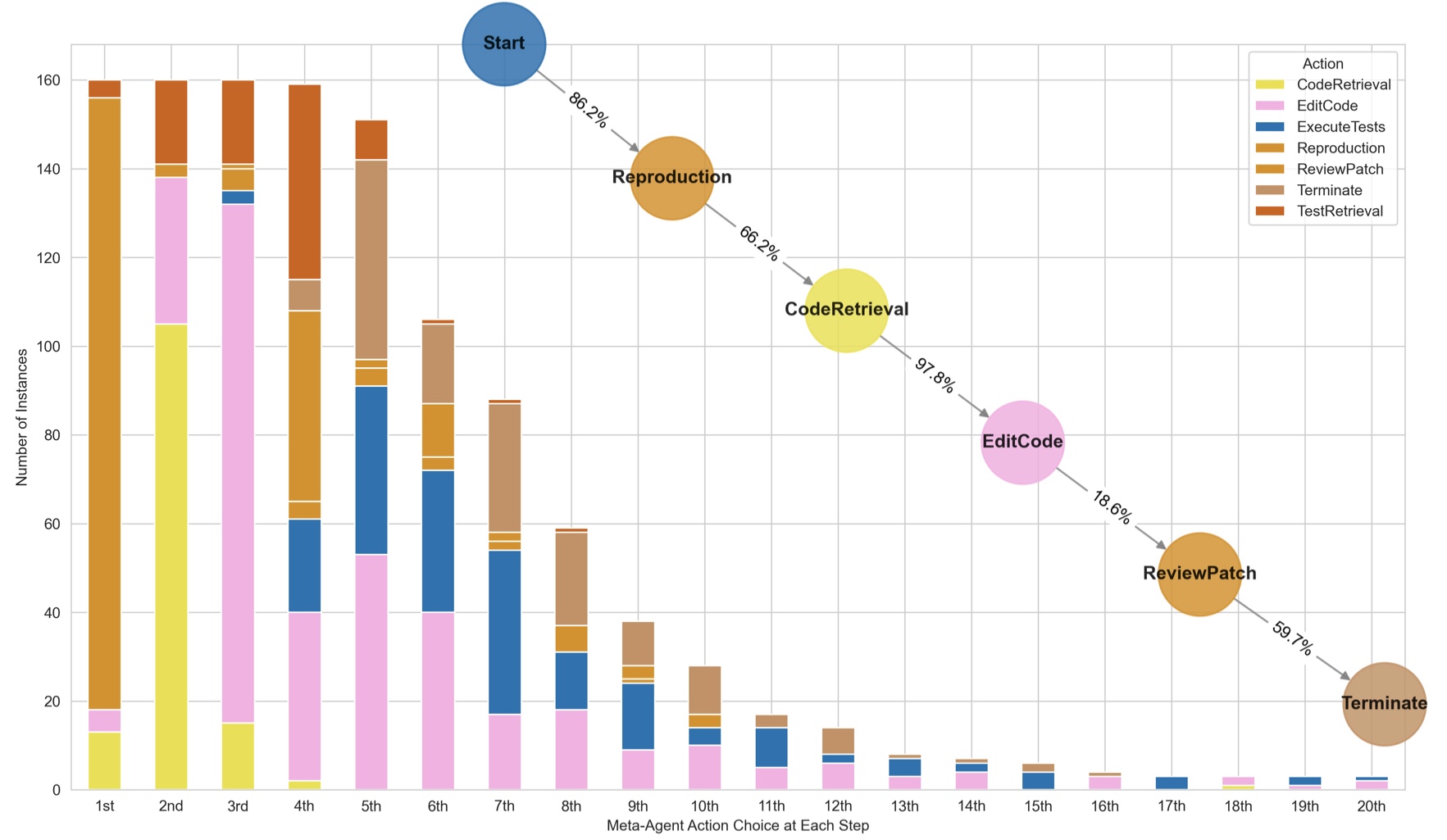}
        \caption{Distribution of SWE Steps -- the most common trajectory begins with reproduction, followed by code retrieval, editing, and review.}
        \label{fig:swe-steps}
    \end{subfigure}
    \hfill    
    \begin{subfigure}{0.49\linewidth}
        \centering
        \includegraphics[width=\linewidth]{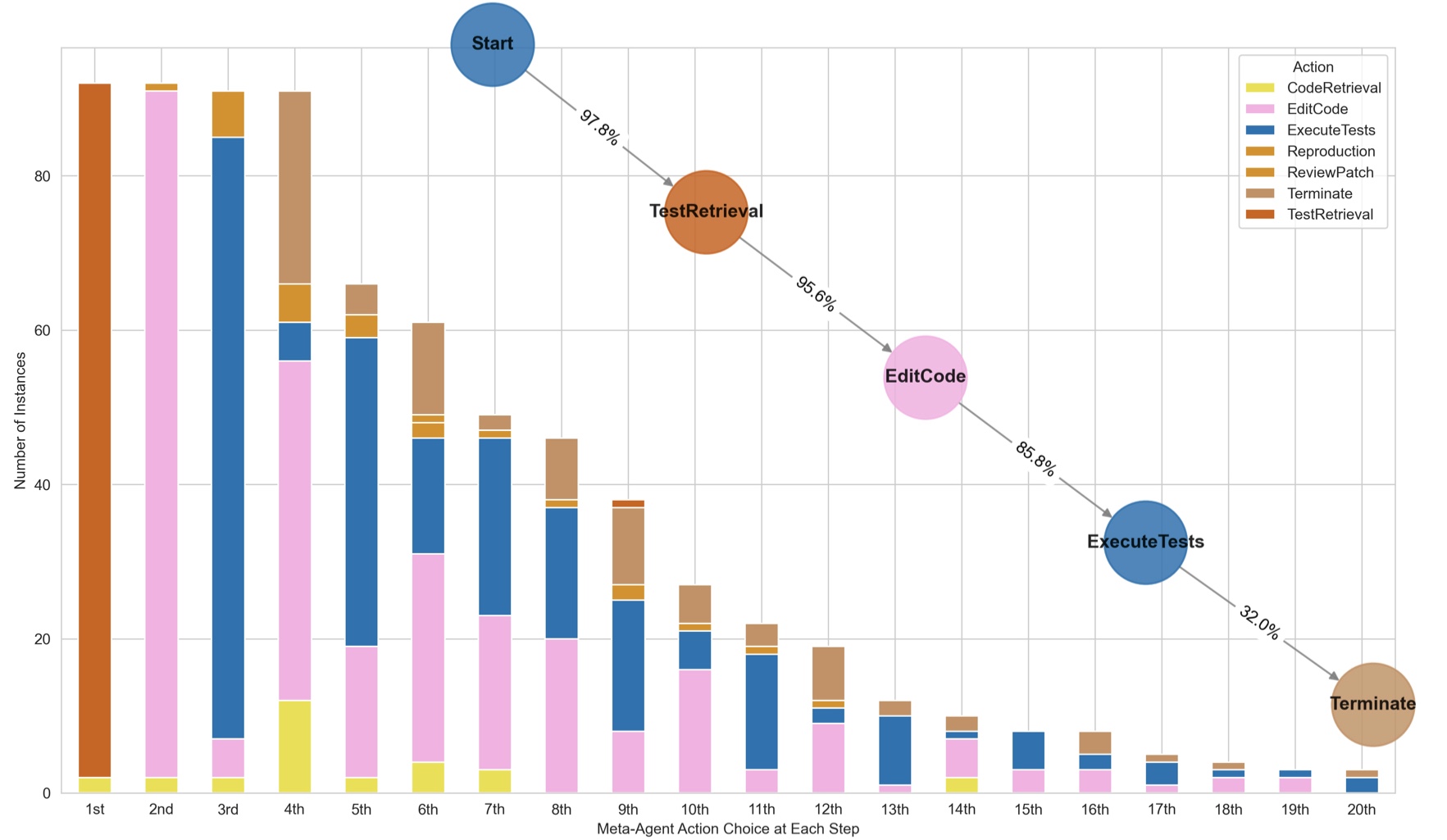}
        \caption{Distribution of SWT Steps - the most common trajectory begins with test-case retrieval, followed by test-editing and execution.}
        \label{fig:swt-steps}
    \end{subfigure}
    \caption{Comparison of SWT and SWE Step Distributions for \useagent on a significant subset of \usebench. Each bar shows how often different actions are invoked at a particular \metaagent step.}
    \label{fig:swt-swe-step-comparison}
\end{figure*}

\begin{table}[t]
    \centering
    \highlighttable
    \footnotesize
    \caption{\highlight{Results of ablation study with \useagent on a statistically significant subset of \usebench, using claude-3-5-sonnet-20241022 as the backend model.}}
    \label{tab:ablation}
    \begin{tabular}{l|ccc}
        \toprule
        \multirow{2}{*}{\textbf{Dataset}}   & \textbf{\useagent} & \textbf{\useagent} & \textbf{\useagent} \\
                   & \textit{original} &  
                   \textit{disable dynamic} & \textit{remove ExecuteTests}  \\
        \midrule
        SWE-Ver  & 49.6\%  & 46.2\% \textbf{(-7\%)}  & 49.6\% \textbf{(-0\%)}  \\
        SWT      & 40.6\%  & 39.1\% \textbf{(-4\%)}  & 36.2\% \textbf{(-11\%)} \\
        REPOCOD  & 8.7\%   & 6.5\% \textbf{(-25\%)}   & 0\%    \textbf{(-100\%)} \\
        REPOTEST & 25.0\%  & 20.0\% \textbf{(-20\%)}  & 17.5\% \textbf{(-30\%)} \\
        % \midrule
        SWETRY   & 13.0\%  & 8.7\% \textbf{(-33\%)}   & 8.7\%   \textbf{(-33\%)} \\
        \midrule
        Overall & 34.9\% & 31.9\% \textbf{(-9\%)} & 31.2\% \textbf{(-11\%)}\\
        \bottomrule
    \end{tabular}
\end{table}

\Cref{fig:swt-swe-step-comparison} shows the different choices of the \metaagent when facing different tasks (we present two for brevity). 
The histograms illustrate the actions taken at each step in the sequence (first, second, etc.).
We observe some clear patterns in the different task types that match our intuition. 
For program repair tasks (i.e. in SWE-Verified), we see in \Cref{fig:swe-steps} that the first step prominently consists of generating a reproduction test, followed by retrieving relevant code context. 
Over the course of agent execution, the \texttt{EditCode} and \texttt{ExecuteTests} become more prominent, introducing and verifying code changes to the software. 
For regression testing (in SWT), \Cref{fig:swt-steps} shows that by-large the first step chosen is \texttt{TestRetrieval}, immediately followed by a (test-)code edit. 
The majority of \useagent trajectories on SWT tasks then consist of alternating test-changes and test-executions.
Overall, the sequence of action invocation follows a certain pattern for each task type as shown in \Cref{fig:swt-swe-step-comparison}, showing that \useagent can self-configure its workflow based on the task type.
Across all benchmarks, \useagent first gathers information (using the retrieval actions or reproducers), 
converging towards alternating \texttt{EditCode} and \texttt{ExecuteTests} until either budgets are exhausted or \texttt{Terminate} is chosen. 
This behavior showcases the capability of emulating concepts like Test-Time Scaling \cite{brown2024large} and command-discovery \cite{bouzenia2024you} through the \metaagent.

\highlight{Table~\ref{tab:ablation} supports these findings through an ablation study. 
The \textit{disable dynamic} column refers to \useagent with the dynamic decision making by MetaAgent disabled. Instead of letting the MetaAgent freely decide which action should be taken as the next step, we predefine the following workflows for the task types:
\begin{itemize}
    \item SWE/SWETRY: \texttt{Reproduction} -> \texttt{CodeRetrieval} -> \texttt{EditCode} -> (\texttt{ReviewPatch} -> \texttt{EditCode})*
    \item SWT: \texttt{TestRetrieval} -> \texttt{CodeRetrieval} -> \texttt{EditCode} -> \\ (\texttt{ExecuteTests} -> \texttt{EditCode})*
    \item REPOCOD: \texttt{EditCode} -> \texttt{TestRetrieval} -> \texttt{EditCode} -> \\ (\texttt{ExecuteTests} -> \texttt{EditCode})*
    \item REPOTEST: \texttt{TestRetrieval} -> \texttt{EditCode} -> (\texttt{ExecuteTests} -> \texttt{EditCode})*
\end{itemize}
where * represents a loop, and the agent can decide when to break out the loop and end the workflow. 
We see that disabling dynamic decision making leads to reductions in efficacy from 4\% to 33\% across tasks.
The MetaAgent in \useagent not only allows for it to handle multiple task types without hurting the efficacy compared to predefined workflows, but without it there would be an efficacy drop.
This efficacy improvement (by the MetaAgent) is largely because the MetaAgent can help to recover from initial sub-optimal decisions (e.g. invoke retrieval actions towards the end of workflow to gather more context). 
}

\highlight{We performed another ablation study to examine the effect of an individual action.
The column \textit{remove ExecuteTests} in Table~\ref{tab:ablation} shows the efficacy of \useagent after removing \texttt{ExecuteTests}, which is one of the actions newly introduced in this work. 
In task types that require repeated test executions to validate and refine solutions (e.g. REPOCOD and REPOTEST), removing \texttt{ExecuteTests} results in significant efficacy drops (-100\% and -30\%).
On the other hand, there is no efficacy drop in SWE-bench tasks, since the MetaAgent could invoke alternative actions such as \texttt{ReviewPatch} to refine solutions.
These results indicate that a single action can enhance effectiveness across multiple task types.
}

\begin{goalbox}{\textbf{Summary: RQ3 - Self Configuration}}
    \useagent is capable of choosing different \textit{action-patterns} for different benchmarks.
    \highlight{The self-configured workflow is as effective as predefined static workflows. 
    In addition, a single action can help improve efficacy in multiple task types.}
\end{goalbox}

\subsection{RQ4: Open Challenges in Agentic Systems}

Based on our investigations, we identify several major challenges in the current unified software engineering agent.
Firstly, when the task requires writing a large amount of code (e.g. feature development tasks), the agent often generates solutions that are ``mostly correct''.
This is reflected on our evaluation on the REPOCOD dataset, where the task is to implement a complex function based on natural language requirement, and the correctness is defined as passing a large number of hidden unit tests.
We observe a number of solutions generated by the agent implement the required feature almost correctly, but fail a few hidden tests due to missing edge cases.
The edge-case behaviors are sometimes not specified clearly in the natural-language task description.
We attribute this challenge in part to the inherent ambiguity of natural language.
A future agent may first resolve the ambiguity in natural language by transforming the task requirements into more ``formal'' artifacts such as specifications and test cases, and then generate solutions based on the formal specification.
Another approach is to design human-agent interaction schemes to clarify ambiguity when needed.

Secondly, the current agent lacks the capability of ``backtracking'' when the agent execution does not yield meaningful results on a given execution path.
For example, when there is a partial patch at a particular program location, the agent is prone to continuously make small edits to the partial patch and execute tests to verify them.
However, the agent may not make good progress with this approach, since it may be impossible to craft a good solution at the location of the partial patch.
This issue is amplified in the SWETRY dataset, in which the task description comes with a partial patch.
To tackle this issue, a future agent should employ a backtracking mechanism to discard unpromising partial solutions, and start over at a previous step.
It could be possible to employ the recent reasoning LLMs to examine the agent execution trace and decide whether to backtrack, since reasoning models have shown backtracking behaviors in their thinking process~\cite{guo2025deepseek}.
Moreover, recent work~\cite{antoniades2024swesearch} on employing search algorithm over agent execution trajectories can help the agent escape from unpromising paths.

Lastly, we still observe patch \textit{overfitting} in tasks such as program repair and regression testing.
Overfitting is a known problem for automated program repair \cite{cacm19}. 
Overfitting happens due to the generated patches passing given tests, but missing actual requirements.
Thus, to avoid overfitting, if the tests are too specific, we need to generalize them. 
Agentic systems show promise for test generation (also reflected in this work), and the \useagent framework allows additional actions, such as test amplification or mutation testing, to be incorporated into the agent.
Artifacts from these actions can be fed back into code generation, thereby reducing overfitting and enhancing trustworthiness. 

\begin{goalbox}{\textbf{Summary: RQ4 - Open Challenges}}
     Challenges include exploring edge-case handling and alternative solutions in feature development, 
     backtracking and discarding poor partial results in repair, 
     and addressing patch overfitting. 
\end{goalbox}

%% file: sections/discussion-related-work-threats.tex
\section{Related Work}
\label{sec:related-work}

\begin{table*}[t]
    \centering
    \highlighttable
    \footnotesize
    \caption{\highlight{Comparing \useagent with LLM agents for software engineering. All these agents take in natural language issues.}}
    \label{tab:related-work}
    \setlength{\tabcolsep}{4pt}
    \begin{tabular}{|p{2.8cm}|p{1.1cm}|p{1.5cm}|p{1.5cm}|p{1.2cm}|p{1.2cm}|p{1.2cm}|p{1.35cm}|p{1.2cm}|p{1.2cm}|}
    \toprule
    & \centering \textbf{\useagent} & \centering \textbf{SWE-Agent} & \centering \textbf{OpenHands CodeAct} & \centering \textbf{AutoCode Rover} & \centering \textbf{Agentless}  & \centering \textbf{MASAI}  & \centering \textbf{Passerine} & \parbox[t]{1.2cm}{\centering \textbf{CodeR}} \\
    \midrule
  
    Is workflow predefined or dynamically constructed? & \centering dynamic & \centering dynamic & \centering dynamic & \centering predefined & \centering predefined & \centering predefined & \centering dynamic & \parbox[t]{1.2cm}{\centering predefined} \\
    \midrule
    
    Handles only program repair or multiple task types? & \centering multiple & \centering multiple & \centering multiple & \centering program repair & \centering program repair & \centering program repair  & \centering program repair & \parbox[t]{1.2cm}{\centering program repair} \\
    \midrule
    
    What is the ``unit of work'' in the agent? & \centering SE action & \centering terminal-level command & \centering terminal-level command & \centering SE action & \centering SE action & \centering SE action  & \centering SE action & \parbox[t]{1.2cm}{\centering SE action} \\
    \midrule
    
    Multi-agent system? & \centering yes & \centering no & \centering no & \centering yes & \centering no & \centering yes  & \centering no & \parbox[t]{1.2cm}{\centering yes} \\
    \bottomrule
    \end{tabular}
\end{table*}

\paragraph{SE - Benchmarks}
Recent Software Engineering Benchmarks fall into two broad categories: 
isolated coding exercises and repository-level tasks. 
Some coding datasets such as CodeXGlue \cite{lu2021codexglue} come without an evaluation metric and form a training corpus; 
others such as HumanEval \cite{chen2021codex}, MBPP \cite{austin2021program}, ClassEval \cite{du2023classeval} or CoderEval \cite{yu2024codereval} provide a challenge accompanied by an evaluation harness (i.e., a test-suite). 
There are already ongoing efforts to provide a meta-benchmark for these tasks \cite{yu2024humaneval,liu2024your}, which commonly target LLMs directly rather than agentic systems. 
For \usebench, we select repository-level tasks to ensure challenges of uniform granularity. 
We opted against coding exercises, as many models have begun to saturate on such benchmarks \cite{sallou2024breaking}.

\paragraph{LLM Agents for Software Engineering} We consider a system \textit{agentic} if the LLM exhibits a degree of autonomy and can interact with the environment through commands or other pre-defined tools. 
Within agentic systems, we distinguish terminal-based agents that interface LLMs directly with a console and minimal tooling (e.g., SWE-Agent \cite{yang2024swe},  Google Jules \cite{google2024gemini2}), from approaches that integrate additional domain knowledge and techniques. 
These additions include specialized tools (such as spectrum-based fault localization in AutoCodeRover \cite{zhang2024autocoderover}), self-reflective capabilities (such as the reviewer agent in SpecRover \cite{ruan2024specrover}), and control logic (such as the finite-state machine in RepairAgent~\cite{bouzenia2024repairagent}). 
In this work, we have focused on AutoCodeRover \cite{zhang2024autocoderover} and OpenHands \cite{wang2024openhands}, as both are widely used open-source projects.
While the Aider project \cite{AIDER2024} covers a well-suited scope involving multiple types of software engineering tasks, its workflow relies on human prompting. 
Passerine \cite{rondon2025evaluating} by Google proposes a program repair framework inspired by SWE-Agent \cite{yang2024swe} and incorporates a structure similar to our Meta-Agent, providing five pre-configured commands (and does not allow free execution of terminal commands, unlike SWE-Agent). 
Our approach differs in scope, as we target a broader range of tasks beyond program repair. 
MASAI \cite{wadhwa2024masai} defines an action space for its central agent similar to our Meta-Agent, but is also limited to program repair.
In contrast, we structure the essential artifacts involved in software debugging via a task state, incorporate a broader set of software engineering actions for orchestration, and apply the agentic framework to multiple task types.
CodeR \cite{chen2024coder} has a Meta-Agent component that pre-defines an execution plan, but does not adapt it dynamically as \useagent does.
\highlight{
Table~\ref{tab:related-work} compares \useagent with existing software engineering agents that take in natural language issues.
\useagent distinguishes itself from the rest by dynamically orchestrating actions (each action is executed by sub-agents) to handle multiple types of software engineering tasks.
}

\begin{figure}[t]
    \centering
    \includegraphics[width=\linewidth]{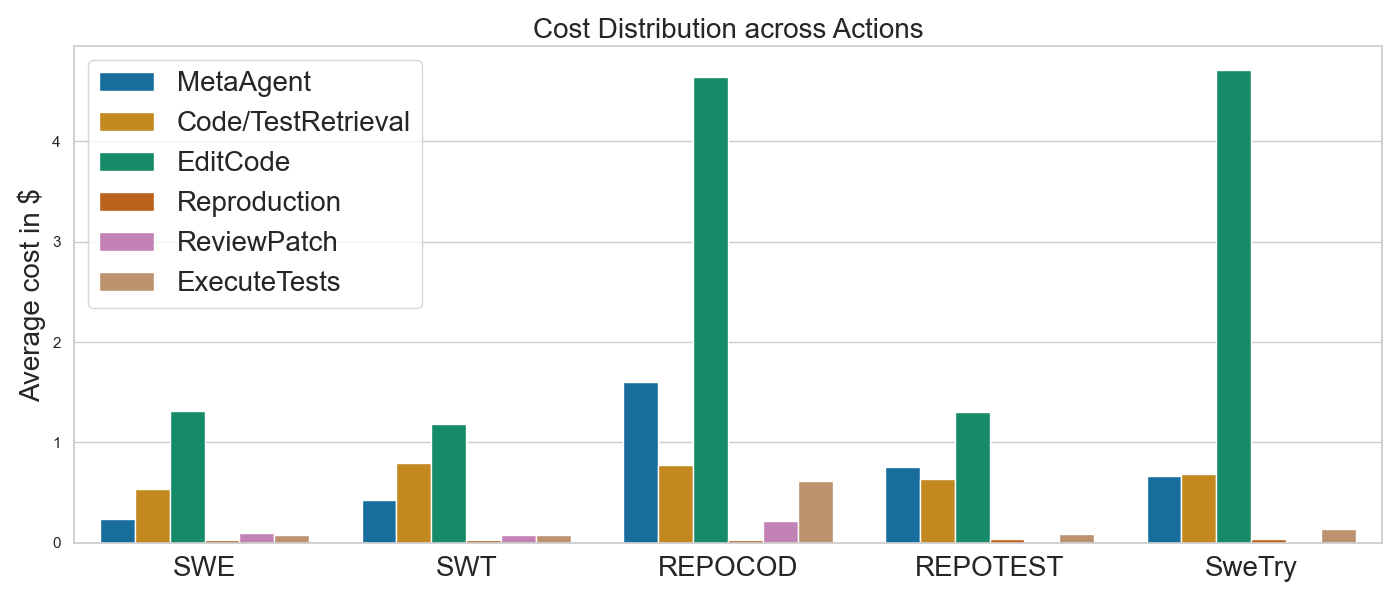}
    \caption{\highlight{Average cost across Actions in \useagent.}}
    \label{fig:useagent-cost-comparison}
\end{figure}

\section{Threats to validity}
\label{sec:threats-to-validity}

\paragraph{Construct Validity} Within \usebench, we approximate testing capabilities using test coverage. 
While test coverage does not guarantee semantic coverage and can sometimes be achieved through various \textit{hacks}, it remains a useful proxy in the absence of a better oracle.
It captures several attributes we aim to evaluate:
code changes must be correctly formatted and target the appropriate program locations. 
Moreover, upon inspection, many of the REPOTEST datapoints contain branches that handle edge-case behavior with errors, where syntactic coverage effectively reflects semantic testing.

\paragraph{Internal Validity: Data Leakage} It is possible that LLMs have been exposed to the projects under evaluation or even datapoints from the source benchmarks \cite{sallou2024breaking}. 
While fully preventing such exposure is beyond our capabilities and the scope of this work, 
we aim to provide an overview of potential memorization by manually inspecting a sample of plausible solutions. 

\section{Discussion}
\label{sec:discussion}

\paragraph*{Cost Analysis}

With Claude 3.5 Sonnet v2 as the backend model, \useagent costs were at an average of 3.65 USD, with notable differences across sub-benchmarks.
Herein SWE scored lowest with 2.20 USD average, and REPOCOD highest at a mean of 7.66 USD. 
The high costs in REPOCOD originate from the iteration of patch generation and test execution.
\highlight{
As shown in \Cref{fig:useagent-cost-comparison}, among all actions, \texttt{EditCode} is the main cost-driver, 
due to the large contexts provided through relevant code and tests.
} 
We calculate the Pearson correlation coefficient for the elapsed time and the costs, and find a strong correlation of $r=0.89$, implying correlation between computational (e.g. test execution) and model (e.g. reasoning) efforts. 
\highlight{When using DeepSeek-V3 as backend model, the average cost of \useagent  is 0.55 USD per datapoint, which is \textasciitilde15\% of the cost with Claude 3.5 Sonnet v2.
Given the efficacy presented in \Cref{tab:efficacy}, open-source models are cost-effective.}

\paragraph*{Perspectives}
We consider the approach of \useagent  to be a promising avenue for solving \textit{any} task that revolves around code as its primary artifact. 
Many tasks such as dependency management or project configuration can be incrementally tackled by  introducing more actions into \useagent.
For \useagent to become truly an \textit{AI Software Engineer}, it must address tasks such as requirements engineering, data visualization, deployment, code review or even a-b-testing. 
The efforts presented in this work are the second step in agentic development towards an AI software engineer, the first being individual agentic systems for single tasks (like AutoCodeRover).
With the \useagent we pave the way for a unified approach to interact autonomously with code regardless of the task.
Going beyond the \useagent will involve studying the cooperative intelligence resulting from multiple USEagents and multiple human developers. 
This might establish the dynamics of future development teams.

\paragraph*{Artifact Availability}
We release the logs and evaluation of USEagent at \url{https://doi.org/10.5281/zenodo.15023393}, the source code of USEbench at \url{https://github.com/nus-apr/USEbench}, and the source code of USEagent at \url{https://github.com/nus-apr/USEagent}.